%% file: main.tex
\documentclass[hidelinks]{article}

\usepackage{arxiv}

\usepackage[utf8]{inputenc} 
\usepackage[T1]{fontenc}    
\usepackage{hyperref}       
\usepackage{url}            
\usepackage{booktabs}       
\usepackage{amsfonts}       
\usepackage{nicefrac}       
\usepackage{microtype}      
\usepackage{lipsum}         
\usepackage{graphicx}
\usepackage[numbers]{natbib}
\usepackage{doi}
\usepackage{todonotes}

\newcommand{\highlights}[1]{
\section*{Research highlights}
}

\usepackage{amssymb}
\usepackage{xspace}
\newcommand{\name}{\emph{\{AnonymizedTool\}}\xspace}
\usepackage{multirow}
\usepackage{listings}

\title{Can We Trust AI-Generated Educational Content? Comparative Analysis of Human and AI-Generated Learning Resources}

\author{ \href{https://orcid.org/0000-0002-5150-9806}
{Paul Denny} \\
	School of Computer Science\\
	The University of Auckland\\
	Auckland, New Zealand \\
   \texttt{paul@cs.auckland.ac.nz} \\
 	\And
	\href{https://orcid.org/0000-0001-8664-6117}{Hassan Khosravi} \\
	Institute for Teaching and Learning Innovation\\
	The University of Queensland\\
	Brisbane, Queensland, Australia \\
	\texttt{h.khosravi@uq.edu.au} \\
	 \AND
	 \href{https://orcid.org/0000-0001-6502-209X}
  {Arto Hellas} \\
	Department of Computer Science\\
	Aalto University\\
	Espoo, Finland \\
	 \texttt{arto.hellas@aalto.fi} \\
	 \AND
	 \href{https://orcid.org/0000-0001-6829-9449}
  {Juho Leinonen} \\ 
	School of Computer Science\\
	The University of Auckland\\
	Auckland, New Zealand \\
	 \texttt{juho.leinonen@auckland.ac.nz} \\
	 \And
	 \href{https://orcid.org/0000-0002-7277-9282}
  {Sami Sarsa} \\ 
	Department of Computer Science\\
	Aalto University\\
	Espoo, Finland \\
	 \texttt{sami.sarsa@aalto.fi} \\
}

\renewcommand{\headeright}{}
\renewcommand{\undertitle}{}
\renewcommand{\shorttitle}{Can We Trust AI-Generated Educational Content?}

\hypersetup{
pdftitle={A template for the arxiv style},
pdfsubject={q-bio.NC, q-bio.QM},
pdfauthor={anon anon},
pdfkeywords={First keyword, Second keyword, More},
}

\begin{document}
\maketitle
\begin{abstract}

\input{article-abstract}

\end{abstract}


\input{article-body}

\section*{Statements on open data, ethics and conflict of interest}

\begin{itemize}

\item {\bf Open data}: All data that was analysed for this work will be made publicly available, along with corresponding Jupyter notebook scripts. 

\item {\bf Ethics statement}: This research was conducted with approval of the \{\emph{Anonymized - details present on title page}\}

\item {\bf Conflict of interest statement}: There are no conflicts of interest to declare.

\end{itemize}

\bibliographystyle{unsrtnat}
\bibliography{refs}  

\end{document}

%% file: article-abstract.tex
As an increasing number of students move to online learning platforms that deliver personalized learning experiences, there is a great need for the production of high-quality educational content.  Large language models (LLMs) appear to offer a promising solution to the rapid creation of learning materials at scale, reducing the burden on instructors.  In this study, we investigated the potential for LLMs to produce learning resources in an introductory programming context, by comparing the quality of the resources generated by an LLM with those created by students as part of a learnersourcing activity. 
Using a blind evaluation, students rated the correctness and helpfulness of resources generated by AI and their peers, after both were initially provided with identical exemplars. 
Our results show that the quality of AI-generated resources, as perceived by students, is equivalent to the quality of resources generated by their peers.  This suggests that AI-generated resources may serve as viable supplementary material in certain contexts.  Resources generated by LLMs tend to closely mirror the given exemplars, whereas student-generated resources exhibit greater variety in terms of content length and specific syntax features used.
The study highlights the need for further research 
exploring different types of learning resources and a broader range of subject areas, and understanding the long-term impact of AI-generated resources on learning outcomes.

%% file: article-body.tex
\section{Introduction}
\label{sec:introduction}

The cornerstone of successful learning communities is ready access to high quality educational resources that enable learners to follow their own learning paths. When paired with a learning management system that keeps track of progress and behavior, large pools of learning resources allow personalized learning experiences that are tailored to match the needs of every student \cite{TORRESKOMPEN2019194}. However, creating and maintaining such large pools of learning resources can take significant effort. One possibility to ease the burden of creating learning resource pools is the use of learnersourcing, where learners are engaged in the collaborative creation and curation of their own learning resources~\cite{kim2015learnersourcing,pirttinen2023lessons}. Although learnersourcing provides ample benefits to students including access to varied educational content and fostering a shared sense of ownership of learning, there are several well-documented issues including low quality and inaccurate resources and challenges around incentivization~\cite{pirttinen2023lessons}. 

The recent emergence of generative AI tools powered by large language models (LLMs) has given rise to a new approach for creating learning resources. Novel content can be created with LLMs, such that the learning objectives and themes of the generated resources can be directed with examples and prompts~\cite{sarsa2022automatic}. 
Despite this great promise, very little research exists that rigorously evaluates the quality of LLM-generated content. In particular, to our knowledge there is no comparative assessment of learning resources that are created by students and by LLMs when both are provided with an identical set of initial examples.  In this work, we prepared six example learning resources suitable for students in an introductory programming course.  We provided these six examples to students to help them get started with a learnersourcing activity, inviting them to create their own similar resources.  We also provided the same six resources to an LLM as few-shot examples, prompting it to generate new ones.  We present the results of a blind evaluation of the quality of both the student-generated and AI-generated resources.  The evaluation involved students rating all learning resources with respect to correctness and helpfulness.  In addition to student perceptions of the learning value of the resources, we are also interested in understanding how the content of the resources differ. The following two research questions guide our work.

\begin{itemize}


\item \textbf{RQ1:}  Given identical priming examples, how do student-generated and AI-generated learning resources differ in terms of overall length and presence of syntax features?

\item \textbf{RQ2:} How do students rate the correctness and helpfulness of learning resources when they are student-generated compared to when they are AI-generated?

\end{itemize}



\section{Related Work}
\label{sec:related}


\subsection{AI in Content Generation}

Large language models (LLMs) have garnered widespread attention in recent years, due to their capability to process and generate text at a very high level of sophistication.  This has resulted in considerable interest in educational settings, where LLMs appear poised to revolutionise teaching and assessment practices.  In particular, LLMs can be applied to the problem of generating high-quality educational resources, including practice problems \cite{sarsa2022automatic}, explanations \cite{leinonen2023comparing}, and textual summaries.  Wang et al. investigated the potential of LLMs for generating questions that could be used in educational settings, and explored a variety of different prompting strategies \cite{wang2022towards}.  They found under certain prompting conditions, the generated questions were indistinguishable to a subject matter expert when compared with human-generated questions.  In similar work, Bulathwela et al. argue that automatic question generation via LLMs will play an important role in scaling online education \cite{bulathwela2023scalable}.  They present a tool which applies an additional pre-training step involving scientific text documents to an existing pre-trained model and then fine-tune it for question generation.  They illustrate that the additional training steps yield higher quality questions, but also note that lack of human evaluation of the generated questions remains an important gap in educational settings. These studies collectively highlight the promising potential of LLMs in automating and personalizing educational tasks, while also emphasizing the need to ensure the quality and relevance of generated content.


However, despite this potential, concerns about the ethical and practical challenges around LLMs have emerged, including concerns around bias in generated content \cite{ferrara2023should}, academic integrity concerns \cite{perkins2023academic}, and the potential impact on educational research and practice \cite{tate2023educational}.  Yan et al. conducted a systematic literature review to identify the practical and ethical challenges associated with LLMs in education, providing recommendations for future research focused on developing practical, ethical, and human-centered innovations \cite{yan2023practical}. Similarly, Chiu et al. performed a systematic review on AI in education, identifying opportunities, challenges, and future research directions across learning, teaching, assessment, and administration domains \cite{CHIU2023100118}. 
One of the key challenges highlighted in this review is the lack of relevant learning resources for personalized and adaptive learning, a challenge which LLMs may be able to help address.

Overall, the literature on LLMs in education presents a complex landscape of both opportunities and challenges \cite{becker2023programming, kasneci2023chatgpt}. The potential of LLMs for automating and personalizing education through content generation is promising, however it is crucial to address ethical, practical, and academic integrity concerns to ensure responsible and effective integration of these technologies into educational systems.




\subsection{Large Language Models in Computing Education}

Across all educational disciplines, there has been significant recent interest in large language models and the challenges and opportunities they present to educators and students \citep{kasneci2023chatgpt}.  Within the field of computing education, where code-generation models face additional constraints on the formation of syntactically and semantically valid outputs, there has also been increasing research interest \citep{denny2023computing, raman2022programming}.  Computing education focused studies have, to date, mostly explored the performance of code generation models on solving typical introductory programming problems \cite{wermelinger2023using, savelka2023generative}. For instance, Finnie-Ansley et al. published a pair of articles that evaluated the performance of the Codex model  on a private database of CS1 and CS2 programming problems from high-stakes end-of-course exams \citep{finnieansley2022robots, finnieansley2023myai}.  In both cases, Codex generated solutions that achieved scores that easily surpassed most students in the course.  A similar study using a public database of programming problems found that Codex was able to produce correct solutions on its first try approximately 50\% of the time, but that with minor modifications to the input prompt this success rate increased to 80\% ~\citep{denny2023conversing}.  In programming-related disciplines such as bioinformatics, where scientists are often not expert programmers, Piccolo et al. showed that ChatGPT could solve three-quarters of basic to moderate level programming tasks on its first attempt, rising to 97\% if allowed up to 7 attempts \cite{piccolo2023bioinformatics}.

Although receiving less attention to date, there has been increasing interest in exploring the capability of AI-based code generation models for producing learning resources. One of the first examples of this work was a study by Sarsa et al. on using Codex for generating programming exercises and code explanations ~\citep{sarsa2022automatic}.  They found the generated code explanations to be both thorough and fairly accurate, with 90\% of the explanations sufficiently explaining all parts of the code, and observed 70\% of the individual line by line explanations to be correct.  However, in this prior work, the evaluation was only performed by experts and the resources were never shown or evaluated by students.  MacNeil et al. also utilized Codex to generate explanations for short code segments, however these were presented to students by placing the explanations next to the corresponding code within in an online course textbook ~\citep{macneil2023experiences}. Their evaluation was limited to around 50 participants, however the students who did access the code explanations found them useful when they chose to engage with them.  Overall, the authors found that the students' engagement was lower than expected and students did not participate in the creation of either the code examples or the accompanying explanations.

\subsection{Learnersourcing}

Learnersourcing is a form of crowdsourcing, where the participants in the crowd are learners~\cite{kim2015learnersourcing}. Typically, learnersourcing is actualized by having students on a course do some task as part of their course work. One common learnersourcing use case is having students create exercises of their own, often in addition to completing traditional exercises \cite{denny2015measuring}. This process usually involves students also reviewing the content produced by their peers. The increasing attention towards learnersourcing has led to new research and methodologies aimed at shaping the development of learnersourcing tools and guiding related research activities. In their study, Khosravi et al. \cite{khosravi2021charting} reflect upon their experience in constructing an adaptive learnersourcing platform, sharing a series of data-driven insights for those in development and research roles. Subsequenty, Singh et al. \cite{singh2022learnersourcing} introduced a theoretical framework to guide the study and design of learnersourcing systems, combining four design questions common in crowdsourcing systems with two questions focused on contributors' skills and learning outcomes. They highlighted the importance of initially addressing the questions of ``what is being done?'' and ``what are contributors learning from the task?'' in any new projects. 


One of the most widely cited examples of learnersourcing is the PeerWise system~\cite{denny2008peerwise}. With PeerWise, students create multiple-choice questions (MCQs), and answer and review MCQs created by their peers. When authoring a new MCQ, a student provides the question text along with answer choices, marking which of the choices is correct. Research exploring the benefits of PeerWise has shown that authoring MCQs can measurably improve student learning~\cite{howe2018peerwise,denny2010peerwise, kelley2019generation}. Other related work has focused on improving student engagement, including exploring the use of various types of gamification, such as leaderboards, points and badges~\cite{denny2018empirical, rogers2021exploring}.

A more recent learnersourcing system is RiPPLE~\cite{khosravi2019ripple}, which is the platform we use in the current study. Similar to PeerWise, RiPPLE supports MCQs, but also includes support for other types of learning resources such as worked examples. 
Recent research involving the RiPPLE platform has explored the potential of learnersourcing as an alternative approach for evaluating the quality of learning resources, finding a strong correlation between student and expert ratings of artefacts while acknowledging variations in students' evaluation skills \cite{abdi2021evaluating}. Research involving the RiPPLE platform has also examined approaches for modelling learners in learnersourcing contexts \cite{lahza2023analytics, abdi2021open}, using AI-assisted methods of improving peer feedback \cite{darvishi2022incorporating,darvishi2022itrust} and the use of explainable AI in education \cite{khosravi2022explainable}.

In the context of programming, a number of domain-specific learnersourcing systems have been developed. For example, both CodeWrite~\cite{denny2011codewrite} and CrowdSorcerer~\cite{pirttinen2018crowdsourcing} support having students create programming exercises, while SQL Trainer supports the creation of SQL exercises~\cite{leinonen2020crowdsourcing}.
In addition to creating exercises, learnersourcing has been used for other tasks. These include, for example, subgoal labels for how-to videos~\cite{weir2015learnersourcing}, personalized hints~\cite{glassman2016learnersourcing}, and pedagogic explanations~\cite{williams2016axis}.

Khosravi et al. recently proposed a framework to guide the development of new learnersourcing approaches that leverage generative AI \cite{khosravi2023learnersourcing}.  Their framework considers several key questions relating to human-AI partnerships that align with different aspects of learnersourcing, for example creating novel content, evaluating content quality, utilising learnersourced contributions and enabling instructors to provide support for students engaging in learnersourcing activities.   In the current work, we focus on the `Create' pillar of the framework proposed by Khosravi et al., by exploring the utility of learning resources generated by large language models.



\subsection{Research Gap}




As discussed in the previous two subsections, prior work has evaluated both learnersourced artefacts and artefacts generated by LLMs. However, to the best of our knowledge, no prior work has evaluated these at the same time, in the same context.  Thus, there currently exists no empirical evidence of the differences in quality between student-created and LLM-created artefacts.

In the present work, we address this research gap by evaluating LLM-created and student-created artefacts in-situ as an authentic learning task, where the artefacts created by students and LLMs are mixed, i.e., the students who review the artefacts do so blind without knowing the source of each artefact. This approach is intended to give an accurate, authentic ranking of the quality of these artefacts.

\section{Methods}
\label{sec:methods}
In this section we describe how artefacts created by students participating in a learnersourcing activity were compared to those generated by an LLM. To begin, we provide an overview of the course context and the learnersourcing activity that was integrated into the course. Next, we introduce the research tool, \name, which was used to conduct the study. Lastly, we outline the study's design, detail the data collected, and discuss the measures used for analysis.


\subsection{Course context}
Our study was conducted in a large first-year programming course taught at the University of Auckland.  This course is compulsory for all first-year students in the engineering programme, uses C as the language of instruction, and had an enrollment of 950 students in 2022.  The course is competitive as, along with all other first-year courses, performance is used to rank students in order to determine entry into limited-place specialisations in the second year.  Typically, fewer than 5\% of enrolled students withdraw from or fail the course.

\subsection{Learnersourcing task}

\begin{figure}[ht]
\centering
\includegraphics[width=\columnwidth]{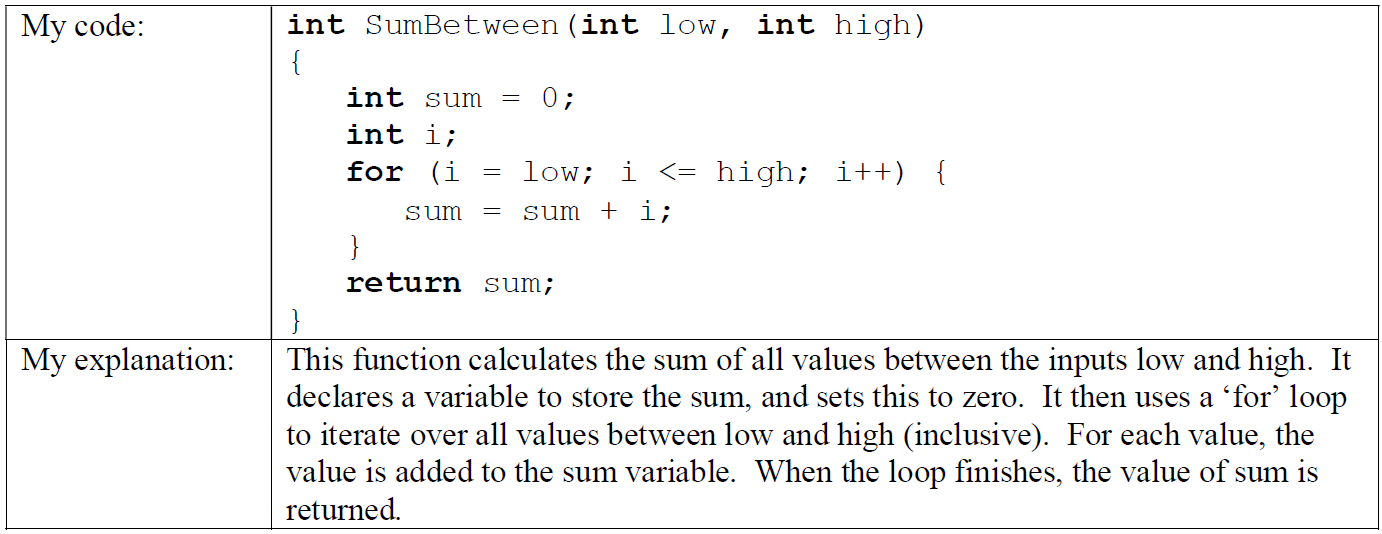}
\caption{A sample ``code example'' artefact  \label{fig:code_example}}
\end{figure}
The learnersourcing task took place during one of the weekly lab sessions.  The lab session is a graded component of the course, and completion contributes 2\% towards each student's grade.  The lab consisted of four short programming tasks, which were completed individually, and the learnersourcing activity was the final task of the lab.  For this task, students were asked to produce a ``code example'' which was defined to consist of two parts, as shown in Figure \ref{fig:code_example}:
(1) a ``My code'' section, where the student had the freedom to write a fragment of code of any length and complexity.  The instructions stated: ``You can provide any code you like for your example. This could be a function you have written, an entire program or even just a short fragment consisting of a few lines of code. It is completely up to you. Ideally, the code you write should highlight one or more concepts that are important in the context of the course. Avoid targeting concepts that are not relevant to our course.''; and (2) a ``My explanation'' section, where the student was required to provide a natural language explanation for the code fragment.  Again, students were mostly free to choose the level of detail contained in their explanation.  At a minimum, they were expected to provide a high-level description of the code.  The following requirements were stated: ``The explanation is simply a short paragraph of text. You should start your explanation with one sentence that describes the overall purpose of the code. After this sentence, you can provide additional details, such as explaining the purpose or function of specific lines of code. Overall, your explanation does not need to be more than a few sentences in length (although feel free to write more)''.

\subsection{Research tool:\name}
For our study, we made use of \name \cite{khosravi2019ripple}, which is an adaptive educational system that develops learning content through learnersourcing. This system offers three types of learning activities: \emph{create}, \emph{evaluate}, and \emph{practice}.  We now briefly describe each of these activities. 

\begin{figure}[ht]
\centering
\includegraphics[width=0.8\columnwidth]{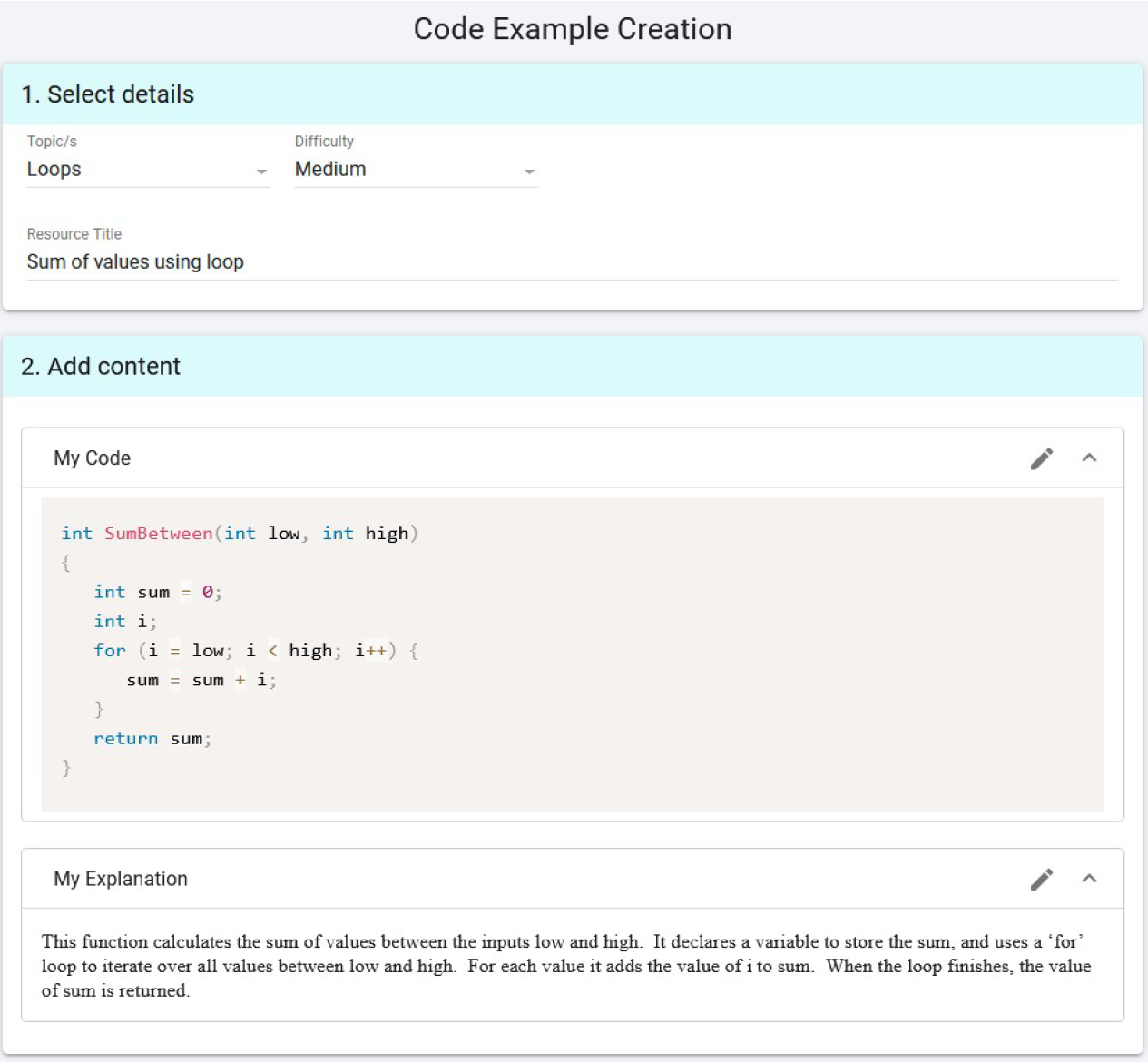}
\caption{The user interface of \name showing the creation process for a new ``code example'' artefact.} \label{fig:RiPPLE_1}
\end{figure}

\textit{Create}. Students in a course develop study resources like MCQs and worked examples. To conduct our study, we introduced a new resource type, ``code example'', as previously described, to the system. To incorporate our new resource type, ``code example,'' we prompted students to select relevant concepts from a list of course topics when creating a new example. They were also asked to indicate the example's level of difficulty (easy, medium, or hard) and provide a brief title. Figure \ref{fig:RiPPLE_1} depicts the interface presented to students for creating a ``code example'' in \name.

\begin{figure}[ht]
\centering
\includegraphics[width=0.8\columnwidth]{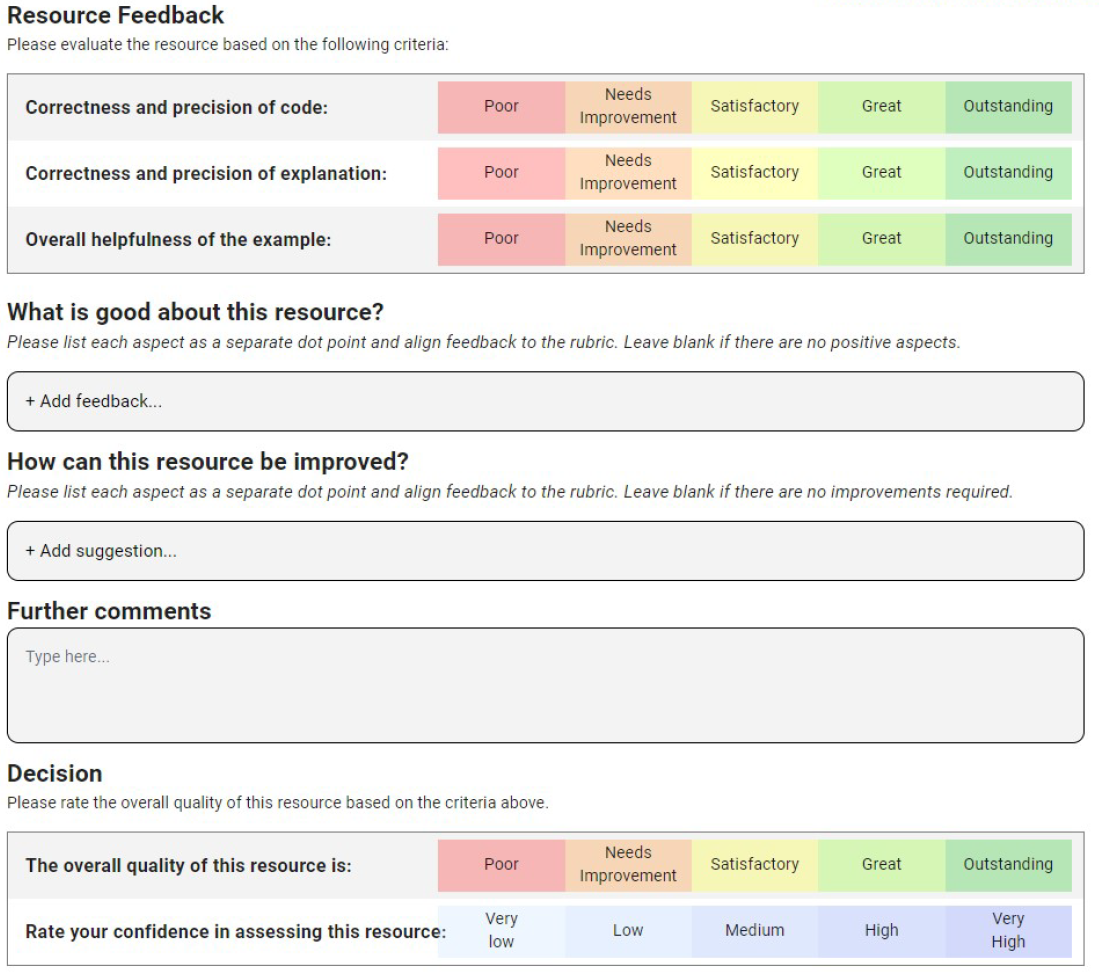}
\caption{The evaluation interface used for reviewing the quality of a ``code example''\label{fig:RiPPLE_2}}
\end{figure}

\textit{Evaluate}. The study resources created by students undergo peer review by multiple peers. Reviewers use a set of criteria to rate the quality of the resource and their confidence in their rating. Additionally, they provide written feedback that justifies their rating and provides constructive feedback on how to improve the resource. Figure~\ref{fig:RiPPLE_2} displays the interface and evaluation criteria for assessing the quality of a code example. Using the submitted peer reviews and algorithms discussed in \cite{darvishi2022incorporating}, \name makes a determination on the quality of the resource. Approved resources are added to a course repository that is accessible to all students while ineffective resources are returned to the author with feedback for resubmission. Instructors can monitor the process with the help of an AI-based spot-checking algorithm \cite{khosravi2021charting}.

\textit{Practice}. The approved resource repository is utilized to provide personalized practice resources to individual students based on their current learning needs.



\subsection{Experimental Design}
To address our research questions, we conducted a study using \name in which students created code examples as part of one activity in a weekly lab session.  Students were provided with six exemplars created by the teaching team to help them craft their own code examples. The code for all six exemplars was provided in the form of a single function definition.  Table \ref{tab:exemplars} shows all six exemplars (code is not included for space reasons).  Three of these six examples (i.e. PrintAverageRainfall, SumBetween and PrintSummary) were published explicitly as part of the description of the lab activity, and the other three were made available as part of a lecture that accompanied the release of the lab resources.   

In a separate process, we used these same six examples to prompt Codex (code-davinci-002; an OpenAI model optimized for code-completion tasks) to generate code examples and their explanations.  We used the default temperature of 0.7 to generate the code examples.  In total, we generated 100 novel code examples using the Codex model.  We began by constructing a prompt with a structure similar to that shown in Listing \ref{lst:codex-prompt}.  The keyword ``EXAMPLE'' was used to denote the beginning of a new example, and each example was divided into a ``Code'' and ``Explanation'' subsection.  Listing \ref{lst:codex-prompt} shows two of the six exemplars, but does not include the other exemplars or the full code and explanations for space reasons.  Of the 100 code examples we generated, 50 were produced using a shorter prompt consisting of a subset of the three exemplars (the three that were published in the description for the lab activity), and the remaining 50 were produced with a longer prompt that included all six exemplars (the three published in the lab description and the three in the accompanying lecture).  The purpose of this split was to facilitate exploration into whether the number of few-shot examples given to the model would measurably affect the quality of the AI-generated resources.  This analysis is outside the scope of the current study and left for future work.  The prompt ended with the keyword ``EXAMPLE'' followed by the keyword ``Code:'' which was intended to prompt the completion model to generate a new example. 
 
\begin{figure}[ht]
\centering
\includegraphics[width=\columnwidth]{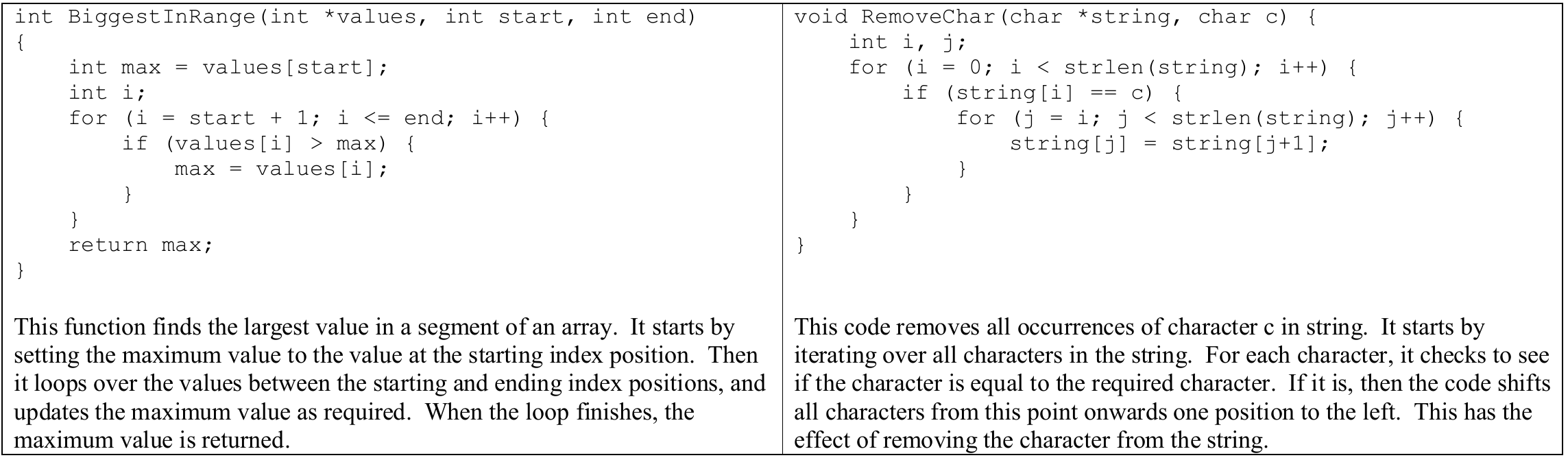}
\caption{Two new ``code examples'' generated as output by Codex \label{fig:codex_outputs}}
\end{figure}

Figure \ref{fig:codex_outputs} illustrates two of the new code examples that were generated by the Codex model.  All 100 code examples generated by Codex were then inserted into \name and appeared in the same pool alongside the student-generated resources.  

\begin{table}[ht]
\caption{Complete list of exemplars that were provided to students and used to formulate the prompt given to the Codex model (in the format shown in Listing \ref{lst:codex-prompt}).  Code is not shown for space reasons. \label{tab:exemplars}}
\centering
\footnotesize{
\begin{tabular}{|l|p{9.5cm}|}
\hline
Function header & Explanation \\
 \hline
void PrintAverageRainfall(int values[])   &  This code is used to get the average rainfall by filtering for the non-negative numbers of the array and adding the numbers up together. Then it calculates the average by dividing the sum by the count of the non-negative numbers. The corresponding result is printed out if the array contains more than one values, and "no rain" is printed out otherwise.  \\ \hline
int SumBetween(int low, int high)   & This function calculates the sum of all values between the inputs low and high. It declares a variable to store the sum, and sets this to zero. It then uses a ‘for’ loop to iterate over all values between low and high (inclusive). For each value, the value is added to the sum variable. When the loop finishes, the value of sum is returned.   \\ \hline
void PrintSummary(int values[])   & The purpose of this code is to print out whether there are more positive numbers or negative numbers in an array. First, variables are declared for counting the number of positive and negative values in the array. Then, there is a loop to iterate through the array. Each time this loop goes through, it will check if the value is positive or negative and count that specific value appropriately. After all numbers are counted correctly, the loop will check if there are more positive or more negative numbers and print out the correct response.   \\ \hline
int IsSuffix(char *suffix, char *word)   &  This function tests whether one string ends with a complete copy of another string, forming a suffix.  It starts by calculating the length of both strings.  Then it iterates over each character in the potential suffix, comparing it to the correpsonding character in the original string such that the lengths line up.  If a character is found not to match, then the function returns 0.  If the loop finishes, then it means all characters match so the function returns 1.  \\ \hline
void RemoveLetterAt(char *word, int position)   & This function removes the character at the specified index position in the string.  It starts looping over the characters in the string beginning from the specified index position, shifting each character one position to the left.  This eventually copies the null character one position to the left, shortening the string as required.   \\ \hline
void ReverseArray(int *values, int left, int right)   &  This function reverses all of the elements in an array that lie between index positions left and right inclusive.  It starts by calculating the length of the segment of the array that lies between the left and right index positions.  It then loops over the first half of the values in this segment, swapping each value with its corresponding value at the other end of the segment.  To swap the values, a temporary variable is used.  On each iteration, the left value increases and the right value decreases.  \\ \hline
\end{tabular}
}
\end{table}


\begin{figure}[ht]
\begin{lstlisting}[caption={Prompt structure for generating Codex code examples and explanations},label=lst:codex-prompt]
EXAMPLE:

Code:
void PrintAverageRainfall(int values[])
{
    ...
}

Explanation:
This code is used...

EXAMPLE:

Code: 
int SumBetween(int low, int high)
{
    ...
}

Explanation:
This function calculates the ...

...

EXAMPLE:

Code:
\end{lstlisting}
\end{figure}

Subsequently, as part of the lab activity, all of the code examples that were produced by students and generated by Codex were peer-reviewed by students using \name's evaluation functionality. Notably, the source of the resources was not shown to students during this review phase, thus the peer evaluation was performed blind. Table~\ref{tab:data} displays the count of the number of resources that were produced by students and by Codex, along with the total number of peer reviews submitted for these resources.

\begin{table}[ht]
\caption{Number of resources created by students and by Codex, and the number of peer evaluations received \label{tab:data}}
\centering
\footnotesize{
\begin{tabular}{|l|l|l|}
\hline
Resource type & \# Resources & \# Evaluation\\
&&\\ \hline
Student-generated   & 886   &4,053   \\ \hline
AI-generated        & 101   & 446   \\ \hline
All                 & 987   & 4,499 \\ \hline
\end{tabular}
}
\end{table}

 \subsection{Metrics and Analysis.}
In this section, we describe the metrics employed and analyses conducted to address our research questions. We employed Mann-Whitney U tests to analyze quantitative data and the Chi-squared test of association for categorical data. A significance criterion of $p < 0.05$ was adopted for assessing statistical significance.

\textbf{RQ1: Comparison of Resources}. Our aim for RQ1 is to assess the comprehensiveness of the resources generated by students and by the AI model. We explore the reserved C keywords used in the ``code'' component of the resources created by Codex, students and instructors (i.e. the six exemplars). Additionally, we gauge the length (character count) of both the ``code'' and ``explanation''  components as an indicator of their comprehensiveness. We present the outcomes of this investigation in Section~\ref{sec:resRQ1}.

\textbf{RQ2: Perceived Quality}. For RQ2, we examined students' perceptions of the quality of student and AI-generated content by analysing their responses to each criteria (code correctness, explanation correctness, and example helpfulness) and their overall decisions regarding the quality of the resources. We present the findings of this investigation in Section~\ref{sec:resRQ2}.

\section{Results}
\label{sec:results}
In this  section we report the results of our investigation into answering the two research questions proposed in Section \ref{sec:introduction}.

\subsection{RQ1: Comprehensiveness of resources created by students vs AI}\label{sec:resRQ1}

Tables~\ref{tab:raw_occurrences},~\ref{tab:resource_occurrences}, and~\ref{tab:1000_characters_occurrences} compare the occurrences for reserved C keywords in the instructor, AI-generated, and student-created code examples. Only keywords that had occurrences for at least two categories (instructor, AI, student) are presented in the tables. In addition to the values shown in the tables, only students had some occurrences for the following keywords: `float', `do', `case', `continue', `short', `unsigned', `true', `struct', `const', `false', `default', `typedef', `auto', `goto', `signed', `static', `switch'.

Table~\ref{tab:raw_occurrences} shows the raw number of occurrences for each keyword in the instructor, the AI-generated, and the student-created code examples. Table~\ref{tab:resource_occurrences} show the average number of times the keyword appeared in a single code example, while Table~\ref{tab:1000_characters_occurrences} shows the average number of occurrences per 1000 characters of code. The values in Table~\ref{tab:resource_occurrences} can be thought of as being normalized with regards to the \emph{number of resources created} while the values in Table~\ref{tab:1000_characters_occurrences} can be thought of as being normalized with regards to \emph{both the number and the length} of the resources created.

There are some interesting observations with respect to Table~\ref{tab:1000_characters_occurrences}. All three groups -- instructors, AI, students -- have different behaviors with regards to keyword use. One interesting observation is that code examples produced by the AI very rarely use the keyword `double' compared to examples produced by instructors and students, with students using `double' the most. The AI uses `while' frequently compared to humans: four times more than students, and slightly under twice as often as instructors. The AI and the instructors use `while' and `for' approximately as often, while students use `while' a lot less often than `for', using `for' three times more often than `while'. Both the AI and students use `else' much less frequently than the instructors. The AI uses `return' almost twice as often as the humans. Students use `if' slightly less compared to instructors and the AI. Interestingly, the number of occurrences of the `int' keyword is different for all three groups: students use it the least, instructors use it more than students, but the AI uses it the most. The difference in the distributions between all three groups (based on the raw number of keyword occurrences) is statistically significant using a Chi-squared test ($p < .001$).


\begin{table}[!ht]
\caption{Comparison of instructor vs AI vs student-generated content based on the raw occurrences of the keywords \label{tab:raw_occurrences}}
    \centering
    \footnotesize{
    \begin{tabular}{|l|l|l|l|l|l|l|l|l|l|l|l|l|}
    \hline
        group & int & if & return & void & else & for & while & char & double & sizeof & break & long \\ \hline
        Instructor & 24 & 7 & 4 & 4 & 4 & 3 & 3 & 3 & 1 & 0 & 0 & 0 \\ \hline
        AI & 378 & 83 & 90 & 40 & 13 & 56 & 54 & 31 & 1 & 5 & 2 & 2 \\ \hline
        Students & 3765 & 1442 & 921 & 706 & 431 & 1000 & 319 & 650 & 366 & 13 & 45 & 8 \\ \hline
    \end{tabular}
    }
\end{table}

\begin{table}[!ht]
\caption{Comparison of instructor vs AI vs student-generated content based on how many times that keyword occurs in a created resource on average \label{tab:resource_occurrences}}
    \centering
    \footnotesize{
    \begin{tabular}{|l|l|l|l|l|l|l|l|l|l|l|l|l|}
    \hline
        group & int & if & return & void & else & for & while & char & double & sizeof & break & long \\ \hline
        Instructor & 4.0 & 1.17 & 0.67 & 0.67 & 0.67 & 0.5 & 0.5 & 0.5 & 0.17 & 0.0 & 0.0 & 0.0 \\ \hline
        AI & 3.74 & 0.82 & 0.89 & 0.4 & 0.13 & 0.55 & 0.53 & 0.31 & 0.01 & 0.05 & 0.02 & 0.02 \\ \hline
        Students & 4.25 & 1.63 & 1.04 & 0.8 & 0.49 & 1.13 & 0.36 & 0.73 & 0.41 & 0.01 & 0.05 & 0.01 \\ \hline
    \end{tabular}
    }
\end{table}

\begin{table}[!ht]
\caption{Comparison of instructor vs AI vs student-generated content based on how often the keyword appears in 1000 characters long code on average \label{tab:1000_characters_occurrences}}
    \centering
    \footnotesize{
    \begin{tabular}{|l|l|l|l|l|l|l|l|l|l|l|l|l|}
    \hline
        group & int & if & return & void & else & for & while & char & double & sizeof & break & long \\ \hline
        Instructor & 10.76 & 3.14 & 1.79 & 1.79 & 1.79 & 1.34 & 1.34 & 1.34 & 0.45 & 0.0 & 0.0 & 0.0 \\ \hline
        AI & 14.86 & 3.26 & 3.54 & 1.57 & 0.51 & 2.2 & 2.12 & 1.22 & 0.04 & 0.2 & 0.08 & 0.08 \\ \hline
        Students & 6.13 & 2.35 & 1.5 & 1.15 & 0.7 & 1.63 & 0.52 & 1.06 & 0.6 & 0.02 & 0.07 & 0.01 \\ \hline
    \end{tabular}
    }
\end{table}

Table~\ref{tab:lengths} presents the lengths of both the code and explanation components of the code examples created by students and the AI. From the table, we can observe that there is a considerable (and statistically significant, Mann-Whitney U = 25123.5, p < .001) difference in the lengths of the code components produced, with the median length being 224 characters for the AI and 377 for the students. For the explanations, the difference is not as large although it is still statistically significant (U = 55834.5, p < .001), and the direction is the opposite, with students producing a median of 303 characters for their explanations compared to 395 characters for the AI.

\begin{table}[ht]
\centering
\caption{Lengths of codes and explanations created by students and AI in characters (rounded to nearest integer). \label{tab:lengths}}
\begin{tabular}{|c|l|l|l|l|l|l|}
\hline
\multirow{2}{*}{}                & Code &                            &                               & \multicolumn{1}{c|}{Explanation} &                            &                               \\ \cline{2-7} 
                                 & Mdn  & \multicolumn{1}{c|}{$\mu$} & \multicolumn{1}{c|}{$\sigma$} & Mdn                              & \multicolumn{1}{c|}{$\mu$} & \multicolumn{1}{c|}{$\sigma$} \\ \hline
Students                         & 377 & 693                     & 1750                        & 303                               & 373                      & 329                         \\ \hline
AI                               & 224   & 252                      & 113                         & 395                               & 405                      & 133                         \\ \hline
\end{tabular}
\end{table}

\subsection{RQ2: Student perceptions of the quality of the resources created by students vs AI}\label{sec:resRQ2}

\begin{table}[ht]
\centering
\caption{Comparison of student vs AI-generated content based on code correctness, explanation correctness, and helpfulness of the generated resource. \label{tab:perception1}}
\begin{tabular}{|c|ccc|ccc|ccc|}
\hline
\multirow{2}{*}{} & \multicolumn{3}{c|}{Code correctness}                                                                              & \multicolumn{3}{c|}{Explanation correctness}                                                                       & \multicolumn{3}{c|}{Helpfulness}                                                                                   \\ \cline{2-10} 
                  & \multicolumn{1}{c|}{Mdn} & \multicolumn{1}{c|}{$\mu$} & \multicolumn{1}{c|}{$\sigma$} & \multicolumn{1}{c|}{Mdn} & \multicolumn{1}{c|}{$\mu$} & \multicolumn{1}{c|}{$\sigma$} & \multicolumn{1}{c|}{Mdn} & \multicolumn{1}{c|}{$\mu$} & \multicolumn{1}{c|}{$\sigma$} \\ \hline
Students          & \multicolumn{1}{c|}{4} & \multicolumn{1}{c|}{4.09}   & \multicolumn{1}{c|}{0.91}     & \multicolumn{1}{c|}{4}   & \multicolumn{1}{c|}{4.01}  & \multicolumn{1}{c|}{0.93}        & \multicolumn{1}{c|}{4}   & \multicolumn{1}{c|}{3.91}  & \multicolumn{1}{c|}{0.94}     \\ \hline
AI                & \multicolumn{1}{c|}{4} & \multicolumn{1}{c|}{4.03}  & \multicolumn{1}{c|}{0.98}     & \multicolumn{1}{c|}{4} & \multicolumn{1}{c|}{4.10}  & \multicolumn{1}{c|}{0.87}      & \multicolumn{1}{c|}{4}   & \multicolumn{1}{c|}{3.89}  & \multicolumn{1}{c|}{0.97}     \\ \hline
\end{tabular}
\end{table}

Table~\ref{tab:perception1} presents student perceptions on the quality of the resources related to correctness of both the code and the explanation and the perceived helpfulness of the resource. From the table, we can observe that while the medians are the same (4 for all evaluated aspects), there are slight differences in the means, although none of the differences are statistically significant (U = 885785.0, p = 0.46 for code correctness; U = 940809.5, p = 0.13 for explanation correctness; U = 894970.0, p = 0.72 for helpfulness).

\begin{table}[ht]
\centering
\caption{Comparison of student vs AI-generated content based on the overall `decision' (overall quality) and confidence in the rater's decision. \label{tab:perception2}}
\begin{tabular}{|c|lll|lll|}
\hline
\multirow{2}{*}{} & \multicolumn{3}{c|}{Decision}                                                                                                                      & \multicolumn{3}{c|}{Confidence}                                                                                                                    \\ \cline{2-7} 
                  & \multicolumn{1}{c|}{Mdn} & \multicolumn{1}{c|}{$\mu$} & \multicolumn{1}{c|}{$\sigma$} & \multicolumn{1}{c|}{Mdn} & \multicolumn{1}{c|}{$\mu$} & \multicolumn{1}{c|}{$\sigma$}\\ \hline
Students          & \multicolumn{1}{l|}{4}   & \multicolumn{1}{l|}{3.94}  & \multicolumn{1}{l|}{0.91}     & \multicolumn{1}{l|}{4}   & \multicolumn{1}{l|}{3.95}  & \multicolumn{1}{l|}{0.78} \\ \cline{1-7}
AI                & \multicolumn{1}{l|}{4}   & \multicolumn{1}{l|}{3.85}  & \multicolumn{1}{l|}{0.97} & \multicolumn{1}{l|}{4}   & \multicolumn{1}{l|}{3.97}  & \multicolumn{1}{l|}{0.81}\\ \hline
\end{tabular}
\end{table}

Table~\ref{tab:perception2} presents student perceptions on the overall quality of the resources, as indicated by their responses to the `decision' field on the evaluation form (this appears towards the bottom of the form as shown in Figure \ref{fig:RiPPLE_2}.  This `decision' criteria measures the overall quality of the resource, and the confidence of the rater in their decision. From the table we can see that while the medians are the same (4 across the board), there are again slight differences in the means. However, the differences are not statistically significant (U = 860373.5, p = 0.07 for decision; U = 922883.5, p = 0.43 for confidence).





\section{Discussion}
\label{sec:discussion}

Results of RQ1 suggest that student-generated learning resources exhibit noticeable variety regarding their overall length and syntax features. In terms of syntax features, the AI, instructors, and students all differ in their usage of specific programming keywords. The divergent use of certain keywords between the AI and students suggests that the AI models may prioritize different programming constructs such as `while' and `return'.  This could have both potential advantages and disadvantages. On the one hand, it can potentially limit the variety of code examples learners are exposed to, impacting their understanding of alternative coding strategies. However, it could also be beneficial if those constructs are indeed more important with respect to the course or applicable to a wider range of problems. Further investigation is needed to determine the impact of this difference on learning outcomes.

The length of AI-generated code appears to be comparable to that of instructors', generally offering more concise functions which might prove more efficient or easier for novices to comprehend. Several factors contribute to the more extended and varied length observed in student-generated content. It seems that students embrace a broader interpretation of the ``code example'' resource type, rather than more closely adhering  to the kinds of examples provided by the instructors. Some students included various components of a full-fledged program, such as include statements, test cases, and comments, in contrast to the instructors' examples which comprised a single function with no comments. These findings point to a greater level of autonomy in students' code writing style, while the AI seems to mirror the given examples and instructions more closely. Explanations generated by the AI are typically similar in length to the instructor's examples and capitalize on the capability of large language models for producing high-quality writing. We observed almost no grammatical or spelling errors in the explanations produced by Codex.  The longer explanations produced by the AI, compared to the student explanations, can be seen as a benefit as it may provide more in-depth coverage of the code. In contrast, student explanations tend to be significantly shorter but exhibit a larger variation in length. This could be attributed to the diverse backgrounds of the student population, which might encompass substantial differences in English language proficiency.

The results of RQ2 demonstrate a comparable perceived quality between student-generated and AI-generated learning resources, in terms of both correctness and helpfulness. This outcome suggests that AI-generated resources could be a viable supplement or alternative to student-generated resources in certain learning contexts. A slightly higher average rating for student-generated code correctness might indicate that students appreciate the broader variety in code examples that may include test cases and comments. On the other hand, the slightly higher mean rating for AI-generated explanations could suggest that the AI's ability to provide detailed and coherent insights is recognized and appreciated by students. Nevertheless, students appear to view student-generated and AI-generated resources largely in similar ways. However, given the close ratings, further research could be beneficial to understand under what conditions one type of resource might be preferred over the other. For example, in our study the explanations that are listed in Table \ref{tab:exemplars} appear quite homogeneous.  Incorporating a wider variety in terms of length and difficulty in the exemplar resources that are provided as prompts to the AI model could potentially stimulate the creation of more diverse learning materials.  
Such an expanded set of examples can provide a richer experience for students and more diverse learning resources. Also, there is a need for studies to examine the long-term impact of the use of AI-generated resources on learning outcomes and to ascertain whether these initial perceptions of the usefulness of AI-generated content hold up over time.

This initial study highlights several potentially fruitful avenues for future work.  As mentioned above, although we did not observe statistically significant differences in the perceived quality of AI-generated and student-generated resources, such difference might become apparent in larger populations of within different learning contexts.  For example, the student cohort in our study are highly incentivised to perform well, and thus are typically well engaged with learning activities.  In situations where students are less engaged, learnersourcing activities are often less successful due to the low quality of student generated content.  In such settings, resources that are automatically generated by AI models and which align well to instructor resources may be particularly valuable. 
The rapid and continuous improvement in LLMs also presents exciting opportunities for the future of AI-generated learning resources.  For example, it may be worthwhile to investigate whether LLMs can be prompted to generate explanations that cater for specific knowledge levels, such as simplifying concepts for students with less strong backgrounds and providing advanced instruction for more capable students.





\section{Conclusion}

In this paper, we explore the potential of employing large language models (LLMs) for generating learning resources in the context of an introductory programming course by contrasting them with resources created by students. Using a blind evaluation we examine the perceived correctness and helpfulness of both AI-generated and student-generated resources.


Our primary findings reveal that the perceived quality of AI-generated resources is largely on par with student-generated resources, though there are notable differences in terms of syntax features and length. This suggests that AI-generated content may serve as a viable supplement or as an alternative to traditional learnersourcing techniques in certain learning contexts. However, the limited variety in AI-generated examples and their close adherence to given prompts raises questions about their adaptability to cater to diverse learning needs and preferences.

In terms of broader lessons, this research highlights the potential for LLMs  to improve the accessibility and scalability of high-quality educational content, while reducing the burden on educators. Future work should focus on investigating the limitations identified in this study, including using a wider variety of input prompts, and seeking to understand the long-term impact of AI-generated resources on learning outcomes.  Our evaluation was also limited to one type of learning resource suitable for introductory programming courses, and future work should explore different kinds of resources in a broad range of subject areas.  The automatic generation of learning resources using LLMs holds great promise for enriching the educational landscape across various domains and skill levels.

